\documentclass[twocolumn]{aastex631}

\usepackage{amsmath}

\defcitealias{marino13}{M13}
\defcitealias{dopita16}{D16}
\defcitealias{pettini04}{PP04}
\defcitealias{curti20a}{C20}

\begin{document}

\title{JWST's TEMPLATES for Star Formation: The First Resolved Gas-Phase Metallicity Maps of Dust-Obscured Star-Forming Galaxies at $z$\,$\sim$\,4}

\correspondingauthor{Jack Birkin}
\email{jbirkin@tamu.edu}

\author[0000-0002-3272-7568]{Jack E. Birkin}
\affiliation{Department of Physics and Astronomy and George P. and Cynthia Woods Mitchell Institute for Fundamental Physics and Astronomy, Texas A\&M University, 4242 TAMU, College Station, TX 77843-4242, US}

\author[0000-0001-6251-4988]{Taylor A. Hutchison}
\altaffiliation{NASA Postdoctoral Fellow}
\affiliation{Observational Cosmology Lab, Code 665, NASA Goddard Space Flight Center, 8800 Greenbelt Rd., Greenbelt, MD 20771, USA}

\author[0000-0003-1815-0114]{Brian Welch}
\affiliation{Department of Astronomy, University of Maryland, College Park, MD 20742, USA}
\affiliation{Observational Cosmology Lab, NASA Goddard Space Flight Center, Greenbelt, MD 20771, USA}
\affiliation{Center for Research and Exploration in Space Science and Technology, NASA/GSFC, Greenbelt, MD 20771}

\author[0000-0003-3256-5615]{Justin~S.~Spilker}
\affiliation{Department of Physics and Astronomy and George P. and Cynthia Woods Mitchell Institute for Fundamental Physics and Astronomy, Texas A\&M University, 4242 TAMU, College Station, TX 77843-4242, US}

\author[0000-0002-6290-3198]{Manuel Aravena}
\affiliation{Instituto de Estudios Astrof\'{\i}sicos, Facultad de Ingenier\'{\i}a y Ciencias, Universidad Diego Portales, Av Ej\'ercito 441, Santiago, Chile}

\author[0000-0003-1074-4807]{Matthew B. Bayliss}
\affiliation{Department of Physics, University of Cincinnati, Cincinnati, OH 45221, USA}

\author[0000-0002-4657-7679]{Jared Cathey}
\affiliation{Department of Astronomy, University of Florida, 211 Bryant Space Sciences Center, Gainesville, FL 32611 USA}

\author{Scott C. Chapman}
\affiliation{Department of Physics and Astronomy, University of British Columbia, 6225 Agricultural Road, Vancouver, V6T 1Z1, Canada}
\affiliation{National Research Council, Herzberg Astronomy and Astrophysics, 5071 West Saanich Road, Victoria, V9E 2E7, Canada}
\affiliation{Department of Physics and Atmospheric Science, Dalhousie University, 6310 Coburg Road, B3H 4R2, Halifax, Canada}

\author[0000-0002-0933-8601]{Anthony H. Gonzalez}
\affiliation{Department of Astronomy, University of Florida, 211 Bryant Space Sciences Center, Gainesville, FL 32611 USA}

\author[0000-0002-7472-7697]{Gayathri Gururajan}
\affiliation{University of Bologna - Department of Physics and  Astronomy “Augusto Righi” (DIFA), Via Gobetti 93/2, I-40129, Bologna, Italy}
\affiliation{INAF - Osservatorio di Astrofisica e Scienza dello Spazio, Via Gobetti 93/3, I-40129, Bologna, Italy}

\author[0000-0003-4073-3236]{Christopher C. Hayward}
\affiliation{Center for Computational Astrophysics, 162 Fifth Avenue, New York, NY, 10010, USA}

\author[0000-0002-3475-7648]{Gourav Khullar}
\affiliation{Department of Physics and Astronomy and PITT PACC, University of Pittsburgh, Pittsburgh, PA 15260, USA}

\author[0000-0001-6505-0293]{Keunho J. Kim}
\affil{Department of Physics, University of Cincinnati, Cincinnati, OH 45221, USA}

\author[0000-0003-3266-2001]{Guillaume Mahler}
\affiliation{Centre for Extragalactic Astronomy, Durham University, South Road, Durham DH1 3LE, UK} 
\affiliation{Institute for Computational Cosmology, Durham University, South Road, Durham DH1 3LE, UK}

\author[0000-0001-6919-1237]{Matthew A. Malkan}
\affiliation{Department of Physics and Astronomy, University of California, Los Angeles, 430 Portola Plaza, Los Angeles, CA 90095, USA}

\author[0000-0002-7064-4309]{Desika Narayanan}
\affiliation{Department of Astronomy, University of Florida, 211 Bryant Space Sciences Center, Gainesville, FL 32611 USA}
\affiliation{Cosmic Dawn Center at the Niels Bohr Institute, University of Copenhagen and DTU-Space, Technical University of Denmark}

\author[0000-0002-4606-4240]{Grace M. Olivier}
\affiliation{Department of Physics and Astronomy and George P. and Cynthia Woods Mitchell Institute for Fundamental Physics and Astronomy, Texas A\&M University, 4242 TAMU, College Station, TX 77843-4242, US}

\author[0000-0001-7946-557X]{Kedar A. Phadke}
\affiliation{Department of Astronomy, University of Illinois, 1002 West Green St., Urbana, IL 61801, USA}
\affiliation{Center for AstroPhysical Surveys, National Center for Supercomputing Applications, 1205 West Clark Street, Urbana, IL 61801, USA}

\author[0000-0001-7477-1586]{Cassie Reuter}
\affiliation{Department of Astronomy, University of Illinois, 1002 West Green St., Urbana, IL 61801, USA}

\author[0000-0002-7627-6551]{Jane R. Rigby}
\affiliation{Observational Cosmology Lab, Code 665, NASA Goddard Space Flight Center, 8800 Greenbelt Rd., Greenbelt, MD 20771, USA}

\author[0000-0002-0786-7307]{J.D.T. Smith}
\affiliation{University of Toledo Department of Physics and Astronomy, Ritter Astrophysical Research Center, Toledo, OH 43606} 

\author[0000-0001-6629-0379]{Manuel Solimano}
\affiliation{Instituto de Estudios Astrof\'{\i}sicos, Facultad de Ingenier\'{\i}a y Ciencias, Universidad Diego Portales, Avenida Ej\'ercito Libertador 441, Santiago, Chile. [C\'odigo Postal 8370191]}

\author[0000-0002-3187-1648]{Nikolaus Sulzenauer}
\affiliation{Max-Planck-Institut f{\"u}r Radioastronomie, Auf dem H{\"u}gel 69 D-53121 Bonn, Germany}

\author[0000-0001-7192-3871]{Joaquin D. Vieira}
\affiliation{Department of Astronomy, University of Illinois, 1002 West Green St., Urbana, IL 61801, USA}
\affiliation{Center for AstroPhysical Surveys, National Center for Supercomputing Applications, 1205 West Clark Street, Urbana, IL 61801, USA}
\affiliation{Department of Physics, University of Illinois, 1110 West Green St., Urbana, IL 61801, USA}

\author[0000-0001-7610-5544]{David Vizgan}
\affiliation{Department of Astronomy, University of Illinois, 1002 West Green St., Urbana, IL 61801, USA}

\author[0000-0003-4678-3939]{Axel Weiss}
\affiliation{Max-Planck-Institut f{\"u}r Radioastronomie, Auf dem H{\"u}gel 69 D-53121 Bonn, Germany}

%
%
\begin{abstract}

We present the first spatially resolved maps of gas-phase metallicity for two dust-obscured star-forming galaxies (DSFGs) at $z$\,$\sim$\,4, from the JWST TEMPLATES Early Release Science program, derived from NIRSpec integral field unit spectroscopy of the H$\alpha$ and [N{\sc ii}] emission lines. Empirical optical line calibrations are used to determine that the sources are globally enriched to near-solar levels. 
While one source shows elevated [N{\sc ii}]/H$\alpha$ ratios and broad H$\alpha$ emission consistent with the presence of an AGN in a $\gtrsim$\,1\,kpc region, we argue that both systems have already undergone significant metal enrichment as a result of their extremely high star formation rates.
Utilizing ALMA rest-frame 380\,$\mu$m continuum and [C{\sc i}]($^3$P$_2$--$^3$P$_1$) line maps we compare the spatial variation of the metallicity and gas-to-dust ratio in the two galaxies, finding the two properties to be anticorrelated on highly resolved spatial scales, consistent with various literature studies of $z$\,$\sim$\,0 galaxies. The data are indicative of the enormous potential of JWST to probe the enrichment of the interstellar medium on $\sim$\,kpc scales in extremely dust-obscured systems at $z$\,$\sim$\,4 and beyond.
\end{abstract}




%
%
\section{Introduction}
\label{sec:intro}

At so-called ``cosmic noon'' ($z$\,$\sim$\,2) and beyond, a significant fraction of the Universe's star formation occurs in dust-obscured star-forming galaxies \citep[DSFGs;][]{casey14,swinbank14, dudzeviciute20,zavala21}. In such systems, the majority of the rest-frame optical and ultraviolet (UV) light associated with young stars is absorbed by dust and reprocessed into the far-infrared. 

Multiwavelength analyses and molecular gas observations have revealed that DSFGs are massive, gas rich and highly star forming, with infrared luminosities $L_{\rm IR}$\,$>$\,10$^{12}L_\odot$, implied star-formation rates (SFRs) of 100--1000\,M$_\odot$\,yr$^{-1}$ \citep[see e.g.][]{swinbank14,spilker15,aravena16,strandet17,dudzeviciute20,reuter20,birkin21} and dense interstellar media \citep[e.g.][]{spilker14,birkin21,rybak22,reuter23}. The observed properties of DSFGs (such as their high IR luminosities) and the fact that sources have been detected up to $z$\,$\sim$\,7 has made them challenging to reproduce in current models of galaxy formation and evolution \citep{dave10,lacey16,mcalpine19,hayward21,bassini22}. Therefore, observations of DSFGs can provide strong constraints on such theories.

The interstellar medium (ISM) plays a critical role in the ongoing processes within DSFGs, such as star formation, supernovae and winds which add, remove, and redistribute metals. Therefore, measuring the gas-phase metallicity and its variation across the galaxy is a powerful indicator of its past evolution \citep[e.g.][]{maiolino19}, for example through scaling relations such as the mass-metallicity relation \citep[MZR; e.g.][]{tremonti04} and fundamental metallicity relation \citep[FMR; e.g.][]{mannucci10, curti20a}. A simple and effective method for estimating metallicity is to measure the relative strengths of the [N{\sc ii}] and H$\alpha$ emission lines \citep[e.g.][]{pettini04,marino13,dopita16,maiolino19}, which suffer similar levels of dust extinction due to their close proximity in wavelength. This method has limitations, such as being highly sensitive to the ionisation parameter and N/O abundance \citep{perez-montero09,dopita16,pilyugin16,peng21}, but the brightness of these lines and their accessibility at high redshifts have made it a popular observable for metallicity \citep[e.g.][]{steidel14,sanders15,gillman22}.

The launch of JWST and its successful commissioning \citep{gardner23,mcelwain23,menzel23,rigby23a,rigby23b} have transformed our ability to study galaxy formation and evolution, and we are now entering a new era for extragalactic astronomy. {JWST's} Near-Infrared Spectrograph \citep[NIRSpec;][]{boeker23} can study high-redshift galaxies at sub-kpc spatial resolution, offering a vast improvement over previous ground-based (and therefore seeing-limited) studies. Additionally, at high redshifts the spectral coverage of NIRSpec encompasses the key emission lines that trace gas-phase metallicity such as H$\alpha$, [N{\sc ii}]6584, [O{\sc iii}]$\lambda$5007,4958, H$\beta$, [S{\sc ii}]$\lambda$6717,31, [O{\sc ii}]$\lambda$3727,29 and [Ne{\sc iii}]$\lambda$3870 \citep[e.g.][]{wuyts14,maiolino19,sanders20}.

At high redshifts, strong gravitational lensing due to massive foreground structures allows us to observe galaxies at higher spatial resolution than would otherwise be possible. This technique is commonly used to study the DSFG population, with modest samples of lensed sources characterized by the {\it Herschel}, {\it Planck}, Atacama Cosmology Telescope (ACT), South Pole Telescope (SPT) \citep[e.g][]{negrello10,vieira13,marsden14,bussmann15,harrington16,spilker16,everett20,kamieneski23a}, and through the ALMA Cluster Lensing Survey \citep[ALCS; e.g.][]{sun22}. Many sources have been detected up to $z$\,$\sim$\,7 \citep[e.g.][]{marrone18,reuter20,endsley23}. 

In this work, we present Director's Discretionary Early Release Science (DD-ERS) NIRSpec IFU observations of two SPT-selected gravitationally lensed DSFGs: SPT0418-17 at $z$\,$=$\,4.2246 and SPT2147-50 at $z$\,$=$\,3.7604 \citep{reuter20}, with which we demonstrate the significant advancements that {JWST} is already providing in our study of this population. The outline of this paper is as follows: in \S\ref{sec:observations} we describe the observations carried out, our data reduction methods, and our analysis of the reduced data. In \S\ref{sec:results} we present the results and discuss their implications. In \S\ref{sec:conclusions} we summarise our findings. Throughout this paper we adopt the cosmology measured by \cite{planck20}, i.e. flat with $\Omega_\mathrm{m}$\,$=$\,0.310 and H$_0$\,$=$\,67.7\,km\,s$^{-1}$\,Mpc$^{-1}$, and a solar metallicity of 12\,$+$\,log(O/H)\,$=$\,8.69 \citep{asplund21}.

\begin{figure*}
\hskip-0.75cm
\includegraphics[width=1.1\linewidth]{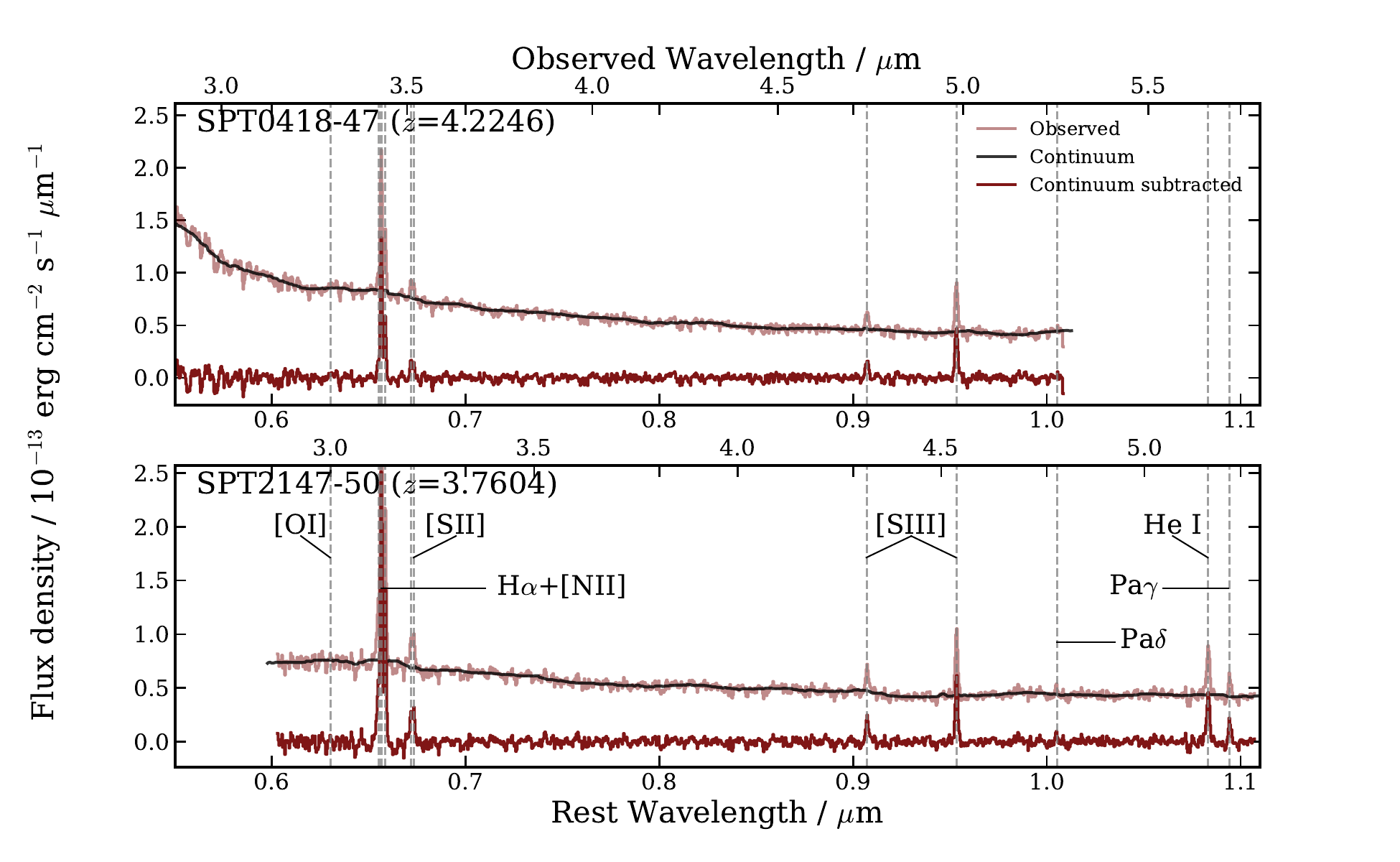}
\caption{
One-dimensional JWST/NIRSpec rest-frame spectra of SPT0418-47 and SPT2147-50, displaying strong detections of the H$\alpha$ and [N{\sc ii}] emission lines, in addition to the [S{\sc iii}] doublet. Both spectra are extracted only from spaxels where H$\alpha$+[N{\sc ii}] is detected with S/N\,$>$\,8. SPT2147-50 displays detections of the He\,I line and Pa\,$\gamma$ lines, and marginal evidence for Pa\,$\delta$ emission. Neither of the sources are detected in [O{\sc i}].
}
\label{fig:1dspectra}
\end{figure*}

\begin{figure*}
\includegraphics[width=0.49\linewidth]{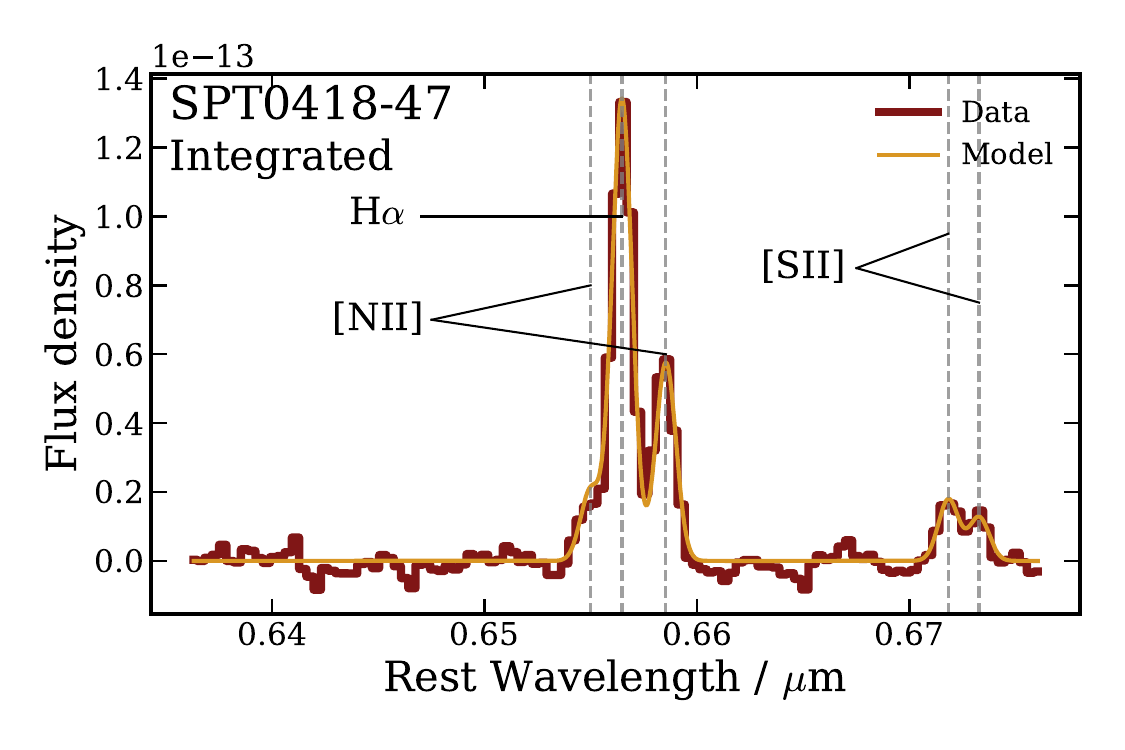}
\includegraphics[width=0.49\linewidth]{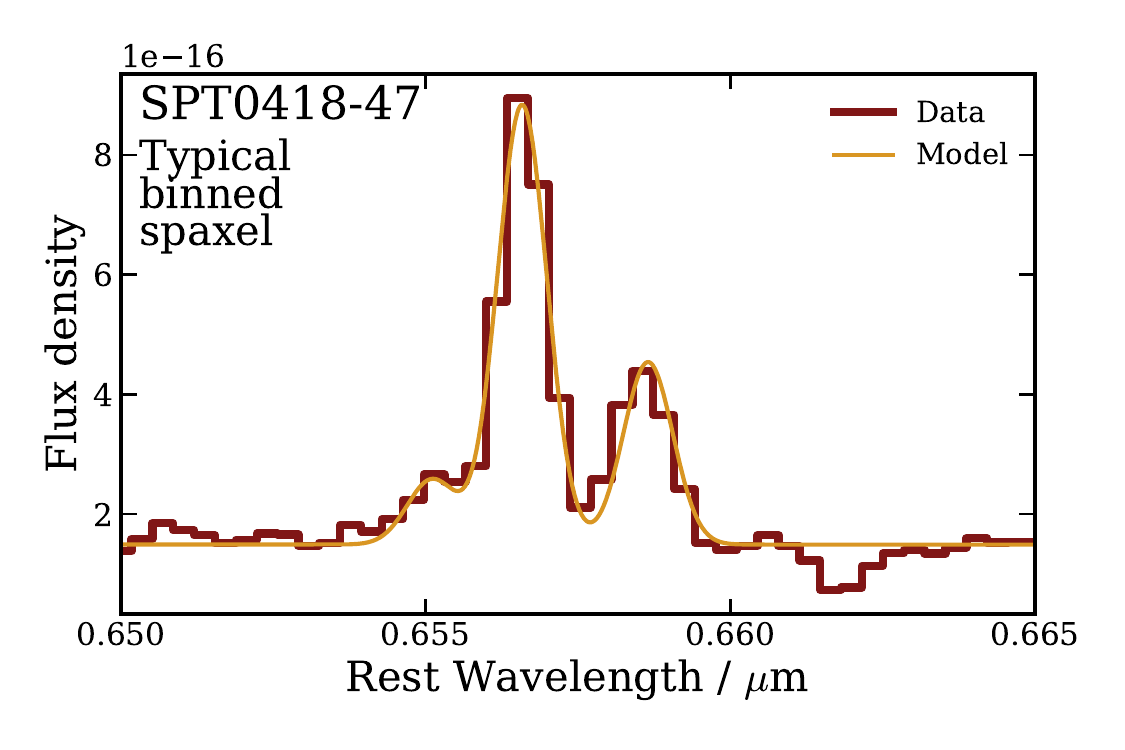}
\includegraphics[width=0.49\linewidth]{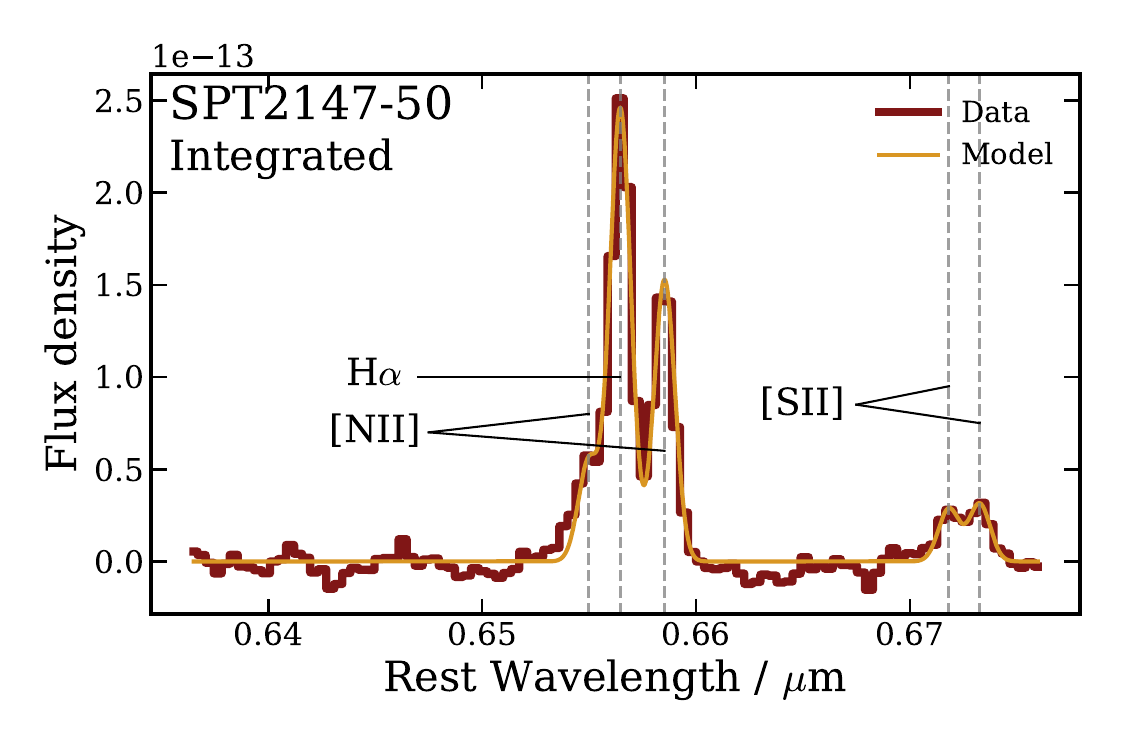}
\includegraphics[width=0.49\linewidth]{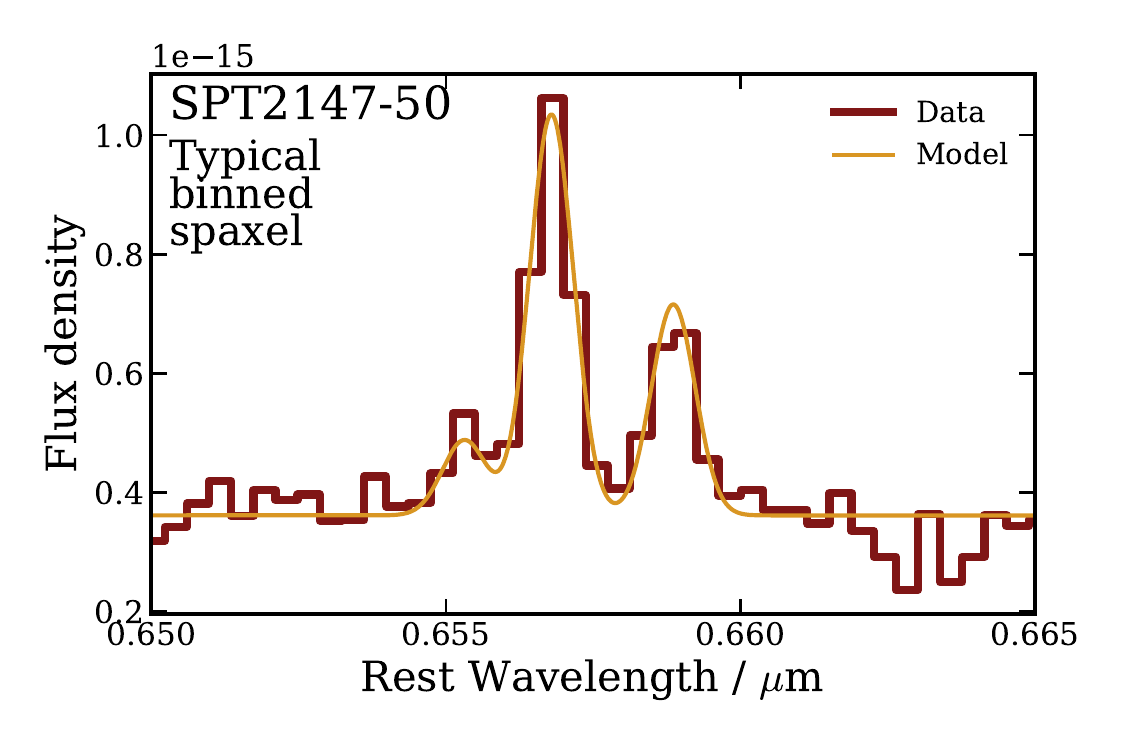}
\caption{Zoom-in 1-D spectra from SPT0418-47 ({\it top}) and SPT2147-50 ({\it bottom}). All flux densities are in units of erg\,cm$^{-2}$\,s$^{-1}$\,$\mu$m$^{-1}$. The left panels show the integrated spectra, and the right panels show a sample typical 3x3 binned spaxel, as used in the spatially resolved fitting. The emission lines are well modelled by Gaussians. We note that the integrated spectra are fit here with [S{\sc ii}] as shown in the plot, but for the binned spaxels we only fit H$\alpha$ and the [N{\sc ii}] doublet, as [S{\sc ii}] is typically too faint to be significantly detected in individual binned spaxels.
}
\label{fig:gaussian_fit}
\end{figure*}

%
%
\section{Observations, data reduction and analysis}
\label{sec:observations}

%
%
\subsection{JWST}
\label{sec:templates}

JWST/NIRSpec observations are drawn from the DD-ERS programme TEMPLATES: Targeting Extremely Magnified Panchromatic Lensed Arcs and Their Extended Star formation (Program 1355; PI: J. Rigby; Co-PI: J. Vieira). From TEMPLATES we utilize NIRSpec IFU G395M/F290LP observations of two SPT-selected DSFGs: SPT0418-47 and SPT2147-50. From spectral energy distribution (SED) fitting it is known that both galaxies are highly massive and highly star-forming \citep[][and see Table \ref{tab:results}]{cathey23}.

The JWST data used in this manuscript can be found in MAST: \dataset[10.17909/fdje-fq59]{http://dx.doi.org/10.17909/fdje-fq59}. We reduce the data with the standard JWST pipeline \citep[Version 1.10.2; ][]{bushouse_howard_2023_7714020} using calibration reference data system (CRDS) context {\tt jwst\_1089.pmap}, with some modifications. Full details on the TEMPLATES data reduction will be provided in Rigby et al. (in prep.), but here we describe how our process differs from the standard pipeline. We modify the outlier detection stage of the pipeline as we found that in some cases the default stage can remove real signal. In our alternate sigma clipping method we first mask the galaxy and clip the background, and then perform the same process on layers of the galaxy of increasing signal-to-noise ratio (S/N). This layering was chosen to ensure that spaxels with weaker emission did not affect the clipping of spaxels with strong emission. We also applied a small offset ($\sim$\,0.5$''$ for both sources) to match the data to ALMA rest-frame 380\,$\mu$m continuum data (see \S\ref{sec:alma}). These astrometric shifts were determined by comparing the coordinates of the central continuum source with {\it HST}/WFC3 F140W imaging \citep{ma15}.

As part of our analysis we also generate NIRSpec cubes with 0.5$''$ resolution (as opposed to the original $\sim$\,0.1$''$ resolution). This is done to ensure consistency with existing ALMA data (see \S\ref{sec:alma}), which we use to derive maps of the dust and gas emission. To generate the 0.5$''$ resolution NIRSpec cubes we convolve the original (0.1$''$ pixel scale) cubes with a circular Gaussian point spread function (PSF) of FWHM 0.49$''$. The target spatial resolution (0.5", and Gaussian by construction) is sufficiently poorer than the intrinsic JWST PSF (0.1") that the wings and spikes in the JWST PSF can be safely ignored.

%
%
\subsection{{\it ALMA}}
\label{sec:alma}

To supplement the data from JWST/NIRSpec, we utilize ALMA data covering the rest-frame 380\,$\mu$m continuum emission and the [C{\sc i}]($^3$P$_2$--$^3$P$_1$) line emission, which will be used in \S\ref{sec:resolved_Z} as proxies for dust mass and gas mass, respectively. For SPT0418-47 we utilize Band 4 data from program 2021.1.00252.S (PI: J. Vieira), which we image using natural weighting and taper to a spatial resolution of 0.5$''$, creating continuum images and CO(7-6)/[C{\sc i}](2--1) cubes. For SPT2147-50 we utilize Band 5 data from program 2018.1.01060.S (PI: J. Vieira) which is processed in the same way, except the imaging was performed using Briggs weighting with a robust parameter of 0.45. This enabled us to achieve the same resolution as the SPT0418-47 data.

\begin{table}[]
    \centering
    \begin{tabular}{c|c|c}\hline\hline
         & SPT0418-47 & SPT2147-50 \\\hline\hline
         R.A. & 04:18:39.67 & 21:47:19.05 \\
         Dec & $-$47:51:52.5 & $-$50:35:53.5\\
         $\mu$ & 29.5\,$\pm$\,1.2 & 6.6\,$\pm$\,0.4 \\
         $\mu$\,SFR / M$_\odot$\,yr$^{-1}$ & 3770\,$\pm$\,545 & 4630\,$\pm$\,230 \\
         $\mu$\,$M_\ast$ / M$_\odot$ & (4.5\,$\pm$\,0.9)$\times10^{11}$ & (4.2\,$\pm$\,1.0)\,$\times$\,10$^{11}$ \\
         $A_V$ & 3.8\,$\pm$\,0.1 & 2.5\,$\pm$\,0.1 \\
         $z$ & 4.2246(4) & 3.7604(2) \\\hline
         \multicolumn{3}{c}{Whole source}\\\hline
         {[N{\sc ii}]/H$\alpha$} & 0.43\,$\pm$\,0.03 & 0.62\,$\pm$\,0.04\\
         {[S{\sc ii}]/H$\alpha$} & 0.23\,$\pm$\,0.01 & 0.24\,$\pm$\,0.01\\
         12\,$+$\,log(O/H)$_{\rm M13}$ & 8.57\,$\pm$\,0.16 & 8.65\,$\pm$\,0.16\\
         $Z_{\rm M13}$ / Z$_\odot$ & 0.8\,$\pm$\,0.3 & 0.9\,$\pm$\,0.3 \\\hline
         \multicolumn{3}{c}{Masking [N{\sc ii}]/H$\alpha$\,$>$\,0.5}\\\hline
         {[N{\sc ii}]/H$\alpha$} & 0.34\,$\pm$\,0.02 & 0.53\,$\pm$\,0.02\\
         {[S{\sc ii}]/H$\alpha$} & 0.21\,$\pm$\,0.01 & 0.23\,$\pm$\,0.01\\
         12\,$+$\,log(O/H)$_{\rm M13}$ & 8.53\,$\pm$\,0.16 & 8.55\,$\pm$\,0.16\\
         $Z_{\rm M13}$ / Z$_\odot$ & 0.7\,$\pm$\,0.3 & 0.7\,$\pm$\,0.3 \\\hline\hline
    \end{tabular}
    \caption{Summary of key properties for both DSFGs studied in this work. Magnification values are taken from \cite{cathey23} for SPT0418-47 and \cite{spilker16} for SPT2147-50. Star-formation rates, stellar masses and dust attenuations are derived from SED fitting with {\sc cigale} \citep[][Phadke et al. in prep.]{cathey23}. Redshifts are taken from \cite{reuter20}. For 12\,$+$\,log(O/H) calibrations, we quote the scatter of the \citetalias{marino13} calibration (0.16\,dex) as the uncertainty, which is much more significant than the uncertainties on our line ratio measurements.}
    \label{tab:results}
\end{table}

%
%
\subsection{Resolved fitting}
\label{sec:resolved_fitting}

To achieve our goal of determining how the metallicity varies across our two targets, we model the H$\alpha$ and [N{\sc ii}] complex as a triple-Gaussian profile, on a pixel-by-pixel basis, tying the wavelengths of the emission and coupling their linewidths, with the H$\alpha$/[N{\sc ii}] flux ratio left as a free parameter. The flux ratio of [N{\sc ii}]$\lambda6584$/[N{\sc ii}]$\lambda6548$ is fixed at 2.8 \citep{osterbrock06}, and we also include a constant continuum component. We attempt to model the emission lines in each pixel, and if we measure a S/N of at least 8 across the H$\alpha$+[N{\sc ii}] complex the pixel is included in the flux maps. Otherwise, the pixel is ignored, and not used in our subsequent analysis. The threshold of S/N\,$=$\,8 was chosen through trial and error, to minimize pixels where the apparent H$\alpha$ emission is largely due to noise while ensuring that the derived flux maps contain enough pixels to study the resolved properties. To boost the S/N we bin each spaxel with its eight neighbouring spaxels. Finally, we visually inspect the resultant H$\alpha$ flux maps along with the fits to each spaxel, and mask by hand any spaxels where the emission does not appear to be coming from the target galaxy. This includes masking emission from a newly detected companion galaxy \citep[][and see \S\ref{sec:line_ratios}]{peng22,cathey23}, which we elect not to study in detail in this work.

\begin{figure*}
\includegraphics[width=\linewidth]{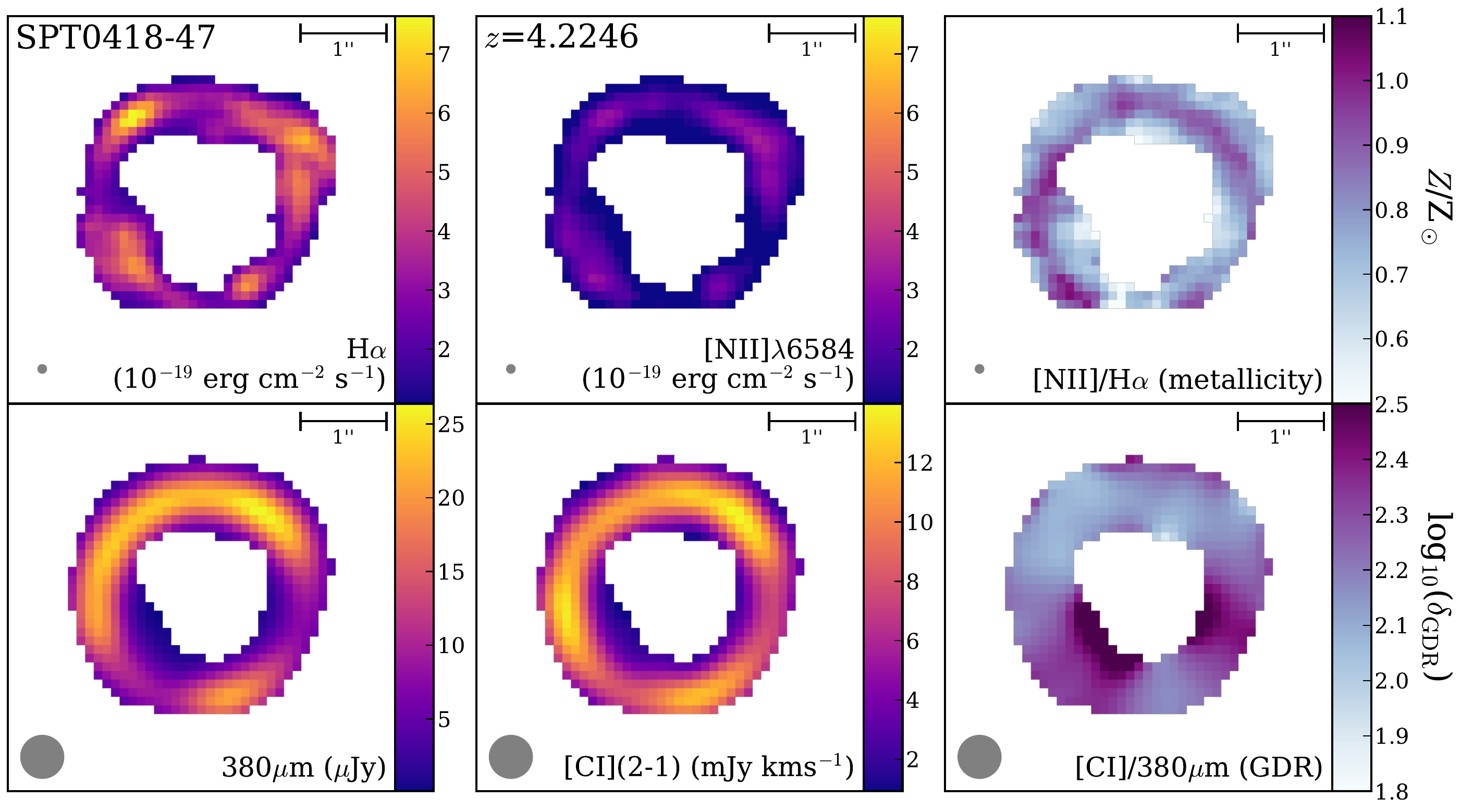}
\caption{{\it Top row:} Resolved H$\alpha$, [N{\sc ii}], and [N{\sc ii}]/H$\alpha$ maps of SPT0418-47 from JWST/NIRSpec ({\it left to right}). The [N{\sc ii}]/H$\alpha$ map is used as a proxy for metallicity. {\it Bottom row:} ALMA rest-frame 380\,$\mu$m continuum and [C{\sc i}] line maps, along with the resultant [C{\sc i}]/380\,$\mu$m map ({\it left to right}). The [C{\sc i}]/380\,$\mu$m map is used as a proxy for the gas-to-dust ratio. The grey circles show the approximate point spread function of the data in each panel, and the horizontal bars in the top right of each panel indicate the scale of 1$''$ ($\sim$\,0.7\,kpc at $z$\,$=$\,4).}
\label{fig:dust_gas_metal_maps_spt0418}
\end{figure*}

\begin{figure*}
\includegraphics[width=\linewidth]{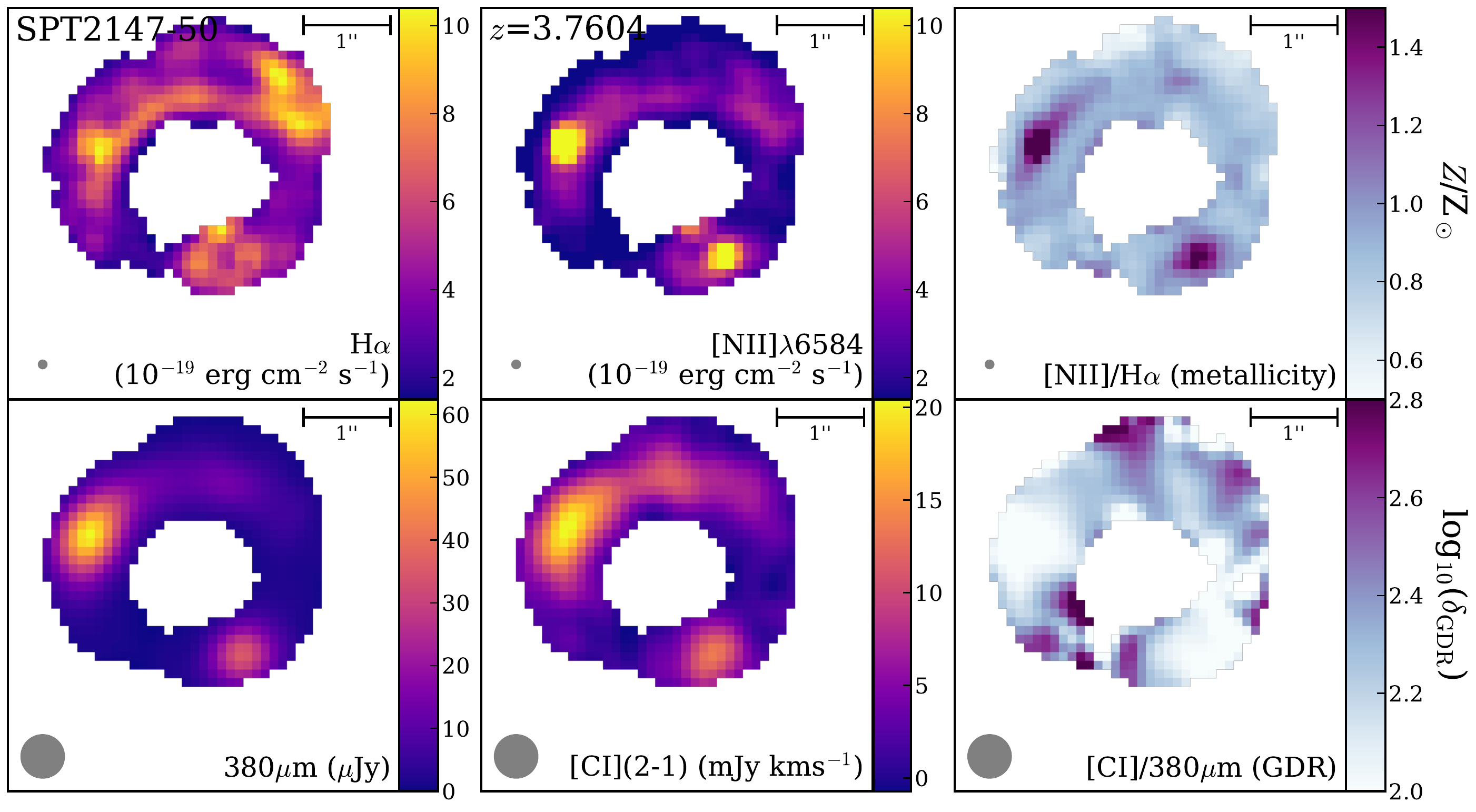}
\caption{Same maps as presented in Fig.~\ref{fig:dust_gas_metal_maps_spt0418}, this time for SPT2147-50. This source displays two regions of very intense [N{\sc ii}] emission which clearly match with regions of strong rest-frame 380\,$\mu$m and [C{\sc i}] emission. Given that [N{\sc ii}]/H$\alpha$\,$>$\,1 in these regions, we are likely detecting emission from an AGN.}
\label{fig:dust_gas_metal_maps_spt2147}
\end{figure*}

%
%
\section{Results and discussion}
\label{sec:results}

Fig.~\ref{fig:1dspectra} shows the 1-D spectra extracted from the two DSFGs in pixels where H$\alpha$ and [N{\sc ii}] are detected, and Fig.~\ref{fig:gaussian_fit} shows the same spectra, but this time zoomed into the H$\alpha$, [N{\sc ii}] and [S{\sc ii}] lines. We show both the integrated spectra, and a sample spaxel (binned with neighbouring spaxels) for each galaxy. The [S{\sc ii}] doublet is shown with a double-Gaussian fit in the integrated spectra as it is clearly detected, but the S/N is insufficient to detect it in individual pixels. Therefore we do not model [S{\sc ii}] in our resolved fitting. The triple Gaussian fit models the H$\alpha$ and [N{\sc ii}] lines well in both cases. The resultant maps of H$\alpha$ and [N{\sc ii}] emission are shown in Figs.~\ref{fig:dust_gas_metal_maps_spt0418} and \ref{fig:dust_gas_metal_maps_spt2147} for SPT0418-47 and SPT2147-50, respectively.
We detect continuum emission across the observed wavelength range in both sources, a combination of stellar continuum from the background DSFGs and light from the bright foreground lenses blurred into the background emission by the JWST PSF. We model the continuum using a running median with a window size large enough so as to avoid removing any flux from the emission lines ($\sim$\,10000\,km\,s$^{-1}$). Both DSFGs display strong H$\alpha$ and [N{\sc ii}]$\lambda6584$ emission, along with the [S{\sc ii}]$\lambda$6717,31 and [S{\sc iii}]$\lambda$9071,9533 doublets. Additionally, in the lower-redshift SPT2147-50 we detect He{\sc i} and Pa$\gamma$ emission, with tentative evidence for Pa$\delta$.

%
%
\subsection{Line ratios and metallicities}
\label{sec:line_ratios}

To derive metallicities for the two DSFGs we first convert the [N{\sc ii}]$\lambda6584$/H$\alpha$ ratio from the line fits to an oxygen abundance 12\,$+$\,log(O/H) using the calibration for N2\,$\equiv$\,log$_{10}$([N{\sc ii}]/H$\alpha$) as derived by \cite{marino13} (hereafter \citetalias{marino13}):
\begin{equation}
\begin{split}
    12+\log({\rm O/H})_{\rm M13} & = 8.743 + 0.462\,{\rm N}2,
\end{split}
\end{equation}
which is derived from a fit to 452 H{\sc ii} regions with $T_e$-based metallicity measurements. The regions considered are valid up to N2\,$=$\,$-$0.2 or [N{\sc ii}]/H$\alpha$\,$\sim$\,0.63. A number of other calibrations were considered, including those proposed by \cite{pettini04} (hereafter \citetalias{pettini04}), \cite{curti20a} (hereafter \citetalias{curti20a}) and \cite{dopita16} (hereafter \citetalias{dopita16}), the latter of which also uses the [S{\sc ii}] doublet. We selected the \citetalias{marino13} calibration as it is considered more reliable at high redshifts and in better agreement with other calibrations than \citetalias{pettini04} \citep[e.g.][]{poetrodjojo21}, and simpler to extrapolate to higher metallicities than \citetalias{curti20a}, who used a fourth-order polynomial fit which is not well constrained in the high-metallicity regime. As an alternative measurement we briefly discuss results using the \citetalias{dopita16} calibration, but as the S/N of our [S{\sc ii}] detections is low we are unable to use this line on a spatially resolved basis. Therefore for consistency we primarily use the \citetalias{marino13} calibration. We note here that while we later study the global mass-metallicity relation and fundamental metallicity relation, our main concern in this work is the spatial \textit{variation} of the metallicity with other properties rather than their absolute values.

As the [N{\sc ii}]/H$\alpha$ ratio is very high ($>$\,0.8) in several regions of both sources, we cannot rule out active galactic nuclei (AGN) as being responsible for this emission. In SPT2147-50 this is supported by very broad line FWHMs ($\sim$\,800\,km\,s$^{-1}$) which correspond with regions of high [N{\sc ii}]/H$\alpha$. This could be evidence for AGN-driven winds in this system. No previous work on these two sources has suggested that either are AGN hosts \citep[e.g.][]{bothwell17,debreuck19}. The same is true for other DSFGs from the same parent sample \citep{ma16,apostolovski19}. Therefore, this is a surprising result, particularly for SPT2147-50 which shows the highest [N{\sc ii}]/H$\alpha$ ratios.

Galaxies are commonly classified as AGN dominated or otherwise using the Baldwin, Phillips and Terlevich \citep[BPT;][]{bpt} diagram \citep[e.g.] []{kewley06}. However, the other BPT diagnostics [O{\sc iii}] and H$\beta$ (neither of which fall within the coverage of our data) are needed to confirm this. Even with these additional diagnostics, the boundary between star-forming galaxies and AGN in the BPT diagram is uncertain at $z$\,$\sim$\,2, let alone at $z$\,$\sim$\,4 where the two DSFGs reside \citep{kewley13}. We therefore also derive metallicities masking out individual pixels with [N{\sc ii}]/H$\alpha$\,$>$\,0.5, which we choose to provide a conservative lower limit on the integrated metallicity.

In the former case, i.e. using all pixels with H$\alpha$ detections, and based on the \citetalias{marino13} calibration, SPT0418-47 has 12\,$+$\,log(O/H)\,$=$\,8.57\,$\pm$\,0.16 and $Z$\,$=$\,0.8\,$\pm$\,0.3\,Z$_\odot$, respectively. SPT2147-50 has 12\,$+$\,log(O/H)\,$=$\,8.65\,$\pm$\,0.16 and $Z$\,$=$\,0.9\,$\pm$\,0.3\,Z$_\odot$, respectively. Adopting a conservative cutoff of [N{\sc ii}]/H$\alpha$\,$>$\,0.5 (which masks $\sim$\,43\,$\%$ and $\sim$\,70\,$\%$ of the pixels in SPT0418-47 and SPT2147-50 respectively) these values change to $Z$\,$=$\,0.7\,$\pm$\,0.3\,Z$_\odot$ for SPT0418-47 and $Z$\,$=$\,0.7\,$\pm$\,0.3\,Z$_\odot$ for SPT2147-50. Interestingly, applying this correction does not significantly change the global metallicity values; both are still a significant fraction of the solar value, and consistent with solar abundances in all cases.

The detections of [S{\sc ii}]$\lambda$6717,31 are significant enough in the integrated spectra of both galaxies (see Fig.~\ref{fig:gaussian_fit}) to provide an independent estimate of the metallicity. Additionally, [S{\sc ii}] is sufficiently close in wavelength to H$\alpha$ and [N{\sc ii}] that any extinction corrections would be negligible. We therefore also apply the \citetalias{dopita16} calibration for H$\alpha$, [N{\sc ii}] and [S{\sc ii}] to estimate 12\,$+$\,log(O/H). Interestingly, this calibration gives significantly higher values of metallicity: $Z$\,$=$\,1.8\,$\pm$\,0.7\,Z$_\odot$ for SPT0418-47 and $Z$\,$=$\,2.7\,$\pm$\,1.0\,Z$_\odot$ for SPT2147-50 (decreasing to 1.5\,$\pm$\,0.6\,Z$_\odot$ and 1.7\,$\pm$\,0.7\,Z$_\odot$ respectively when applying our AGN masking). The \citetalias{dopita16} calibration has been claimed to have a reduced dependence on the ionisation parameter when compared to N2, and therefore this is further evidence that the two DSFGs are highly enriched with metals, possibly even to solar or super-solar levels.

We note here that the two [S{\sc iii}] lines that we detect here can also be used to trace the ionization parameter and metallicity. There remains uncertainty over the reliability of NIRSpec flux calibrations as a function of wavelength, and so we prefer to only use ratios between emission lines that are close in wavelength. Additionally, we would require the application of extinction corrections to our data, given the large wavelength difference between [S{\sc iii}] and the other lines. As we do not have measurements of the H$\beta$ flux, and therefore the Balmer decrement, our extinction corrections would be highly uncertain. Therefore we have elected not to use the [S{\sc iii}] lines in our analysis.

As a side note, a companion source to SPT0418-47 (SPT0418-47B) has been detected by \cite{peng22} and \cite{cathey23}. While we are not focused on the nature of this source here, we note that we measure a metallicity of $\sim$\,0.6--0.7\,Z$_\odot$ (applying the \citetalias{marino13} calibration), which we find to be consistent with \cite{peng22}. Therefore the smaller merging companion to SPT0418-47 is a slightly less metal-rich system.

A potential source of uncertainty in our  metallicity estimates may arise from diffuse ionized gas (DIG), as the resolution of our observations is much greater than the scale of individual H{\sc ii} regions. \cite{valeasari19} investigated this effect for star-forming galaxies in the MaNGA survey and found the diffuse component to have a $\sim$\,0.1\,dex contribution to metallicity estimates derived using the [N{\sc ii}]/H$\alpha$ index, at the high-metallicity end. They also found the DIG to have a more prominent effect on the mass-metallicity relation at the high stellar mass end, where our sources (particularly SPT2147-50) lie. Similar findings were also made by \cite{poetrodjojo19}. A correction for this potential effect is beyond the scope of this work, however.

In general, due to the systematic uncertainties in different calibrations and the fact that these are derived for galaxies at much lower redshifts than the two SPT DSFGs, it is highly challenging to confidently constrain their global metallicities. However, even while being conservative over the possibility of AGN emission in the two DSFGs, we can confidently say that both are enriched to near-solar levels, likely as a result of their very high SFRs.

\begin{figure}
\includegraphics[width=\linewidth]{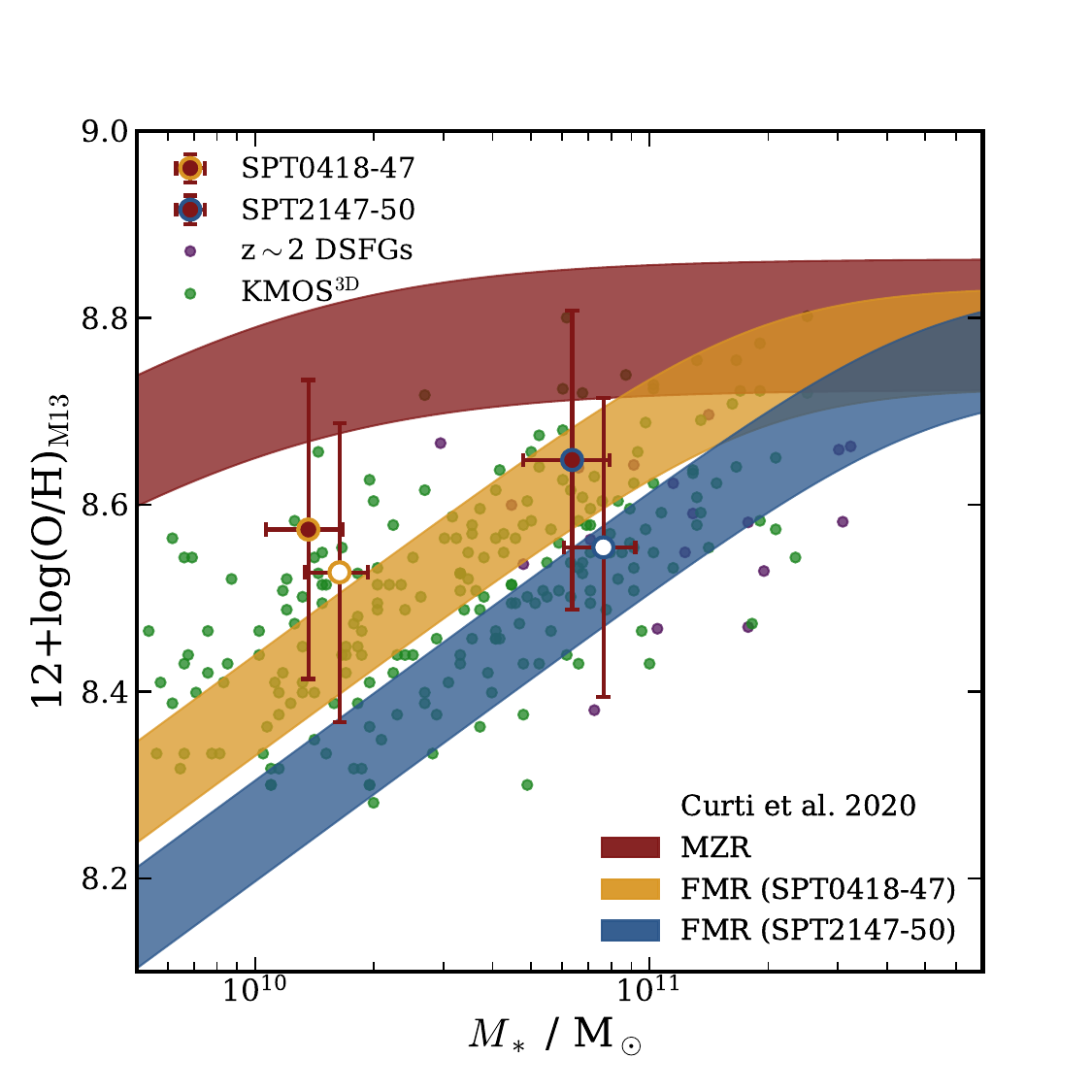}
\caption{Oxygen abundance 12\,$+$\,log(O/H) estimated from the [N{\sc ii}]/H$\alpha$ ratio, using the \citetalias{marino13} calibration (see \S\ref{sec:line_ratios}), versus stellar mass estimated from SED fitting. As comparison samples we include galaxies from the KMOS$^{\rm 3D}$ survey \citep{wisnioski15}, and $z$\,$\sim$\,1.5--2.5 DSFGs also observed with KMOS \citep{birkin_thesis}. From \cite{curti20a} we display the mass-metallicity relation (MZR, red shaded) and the fundamental metallicity relation (FMR) for the SFRs of the two galaxies (yellow for SPT0418-47, blue for SPT2147-50). The open points show how far down the points would move if we were to conservatively exclude pixels with [N{\sc ii}]/H$\alpha$\,$>$\,0.5 to account for AGN (the points are also shifted horizontally for visual clarity). Both DSFGs are consistent with the mass-metallicity relation, and are also consistent with the upper end of the fundamental metallicity relation, particularly when we account for potential AGN emission.
}
\label{fig:mzr}
\end{figure}

%
%
\subsection{Mass-metallicity relation and the fundamental metallicity relation}
\label{sec:mzr}

In general, galaxies appear to display a correlation between gas-phase metallicity and stellar mass, the so-called Mass-Metallicity Relation \citep[MZR; e.g.][]{tremonti04}. In addition, a ``three-dimensional'' relation between stellar mass, metallicity and SFR has been suggested and named the Fundamental Metallicity Relation \citep[FMR;][]{mannucci10}. In the FMR, low-mass galaxies with higher SFRs typically contain a lower proportion of metals, and in high-mass galaxies the metallicity saturates and becomes independent of SFR (as in the mass-metallicity relation). \cite{mannucci10} found this relation to hold constant with low scatter up to $z$\,$\sim$\,2.5 \citep[also see e.g.][]{lara-lopez10,troncoso14,curti20a}, attributing the result to the dominance and consistency of smooth secular processes at low to intermediate redshifts, with this equilibrium being reached at least as early as $z$\,$\sim$\,2.5.

At higher redshifts the picture is less clear. \cite{mannucci10} showed that galaxies at $z$\,$\sim$\,2.5 are offset from the FMR, specifically around 0.6 dex towards lower metallicities, which may imply that smooth secular processes are less prominent in higher redshift sources. To test this for the two $z$\,$\sim$\,4 DSFGs, in Fig.~\ref{fig:mzr} we show the derived oxygen abundances 12\,$+$\,log(O/H) using the \citetalias{curti20a} calibration versus stellar masses derived from SED fitting with {\sc cigale} \citep[][Phadke et al. in prep.]{cathey23}. The open points indicate the oxygen abundances derived from masking emission in pixels with [N{\sc ii}]/H$\alpha$\,$>$\,0.5, and are shifted horizontally for visual clarity.

For comparison with other similarly selected sources, we also include in Fig.~\ref{fig:mzr} $z$\,$\sim$\,2 DSFGs with [N{\sc ii}]/H$\alpha$ measurements from the $K$-band Multi-Object Spectrograph \citep[KMOS;][]{birkin_thesis}, along with $z$\,$\sim$\,1.3--2.6 galaxies from the KMOS$^{\rm 3D}$ survey \citep{wisnioski15}, the latter of which is generally comprised of galaxies with lower SFRs than those studied in this work. The two DSFGs presented in this work are consistent with the majority of the comparison sources, albeit generally on the more metal-rich end.

From \citetalias{curti20a} we show the mass-metallicity relation for local galaxies, along with the FMR for the SFRs of the two galaxies (see Table \ref{tab:results}), including the scatter in both cases. Both galaxies are consistent with the local mass-metallicity relation within the uncertainties --- they are high-mass galaxies, at which point the metallicity is expected to saturate at around 12\,$+$\,log(O/H)\,$\sim$\,8.7--8.8. According to the FMR they are both marginally more metal rich than expected, although generally consistent within the large uncertainties, especially when we account for potential AGN emission.

\cite{mannucci10} claim that the relatively small scatter in the FMR up to $z$\,$\sim$\,2.5 implies constant relative significance of star formation, infall and outflows of gas. The two galaxies presented in this paper appear to have metallicities consistent with the FMR and could therefore fit into this picture. However, SPT0418-47 seems to be interacting with a companion \citep[][]{peng22,cathey23}, which may be expected to drive it away from the FMR. Studies using both observations and simulations have shown that in the early stages of a merger the gas is diluted, thus lowering the metallicity. Subsequently, the metallicity is increased as the merger proceeds, when the star-formation rate is enhanced \citep[e.g.][]{rupke08,montuori10,rupke10}. It is therefore interesting that SPT0418-47 remains consistent with the FMR. We choose not to overinterpret this result given the large uncertainties on the oxygen abundances.

We note that we use the relations from \citetalias{curti20a} as \cite{marino13} do not derive either the FMR or the MZR from their calibration. However, we note that we would expect both to move down if these relations were available for the \citetalias{marino13} calibration, and if anything this would place the two DSFGs $\sim$\,0.2\,dex further above the FMR than we see in Fig.~\ref{fig:mzr}. This would put both sources firmly above the FMR.

An important consideration to make is whether or not the line calibrations we assume are applicable at $z$\,$\sim$\,4. There are few $T_e$ method-based metallicity measurements calibrated to H$\alpha$ and [N{\sc ii}] at this epoch \citep[{although some work has been done with H$\beta$ and [O{\sc iii}]; e.g.}][]{curti23,trump23}, hence our reliance on local calibrations, and it is possible DSFGs exhibit different relative abundances than nearby galaxies. For instance, a top-heavy initial mass function, as has been suggested as necessary to reconcile DSFG number counts with theory \citep[e.g.][]{baugh05}, and therefore a large number of high-mass stars, would lead to a change in the relative abundance of nitrogen, oxygen and hydrogen. A lack of observational data limits our ability to draw the correct conclusion, however some work on this topic has already been done with NIRSpec. For example, for $z$\,$=$\,2.18 galaxies \cite{sanders23} have found evidence that the local \citetalias{curti20a} N2 calibration for metallicity is biased $\sim$\,0.5\,dex high when compared to direct measurements from auroral lines. This bias may be as significant, or potentially even worse, at $z$\,$\sim$\,4.

A further consideration is that if SPT2147-50 does in fact host an AGN that is contributing significantly to the infrared emission, then this could result in a larger estimate of the SFR from SED fitting than the true value. This would in turn mean that the expected metallicity from the FMR is actually lower than it should be. Additionally, the SFRs used in the \citetalias{curti20a} scaling relations are derived from the H$\alpha$ luminosities, scaled to a total SFR. Given the high levels of dust extinction in the two galaxies (see Table \ref{tab:results}), we do not derive H$\alpha$-based SFRs in this work, as the corrections are likely to be significant and uncertain. We instead use total SFRs derived from {\sc cigale} SED fitting, which uses the parameterization from \cite{boquien19}, with the star-formation histories modelled as an exponential decay ($\tau_{\rm burst}$\,$=$\,1\,Gyr) including an additional burst component using the stellar population model from \cite{bruzual03}.

\begin{figure*}
\hspace{-2cm}
\includegraphics[width=1.2\linewidth]{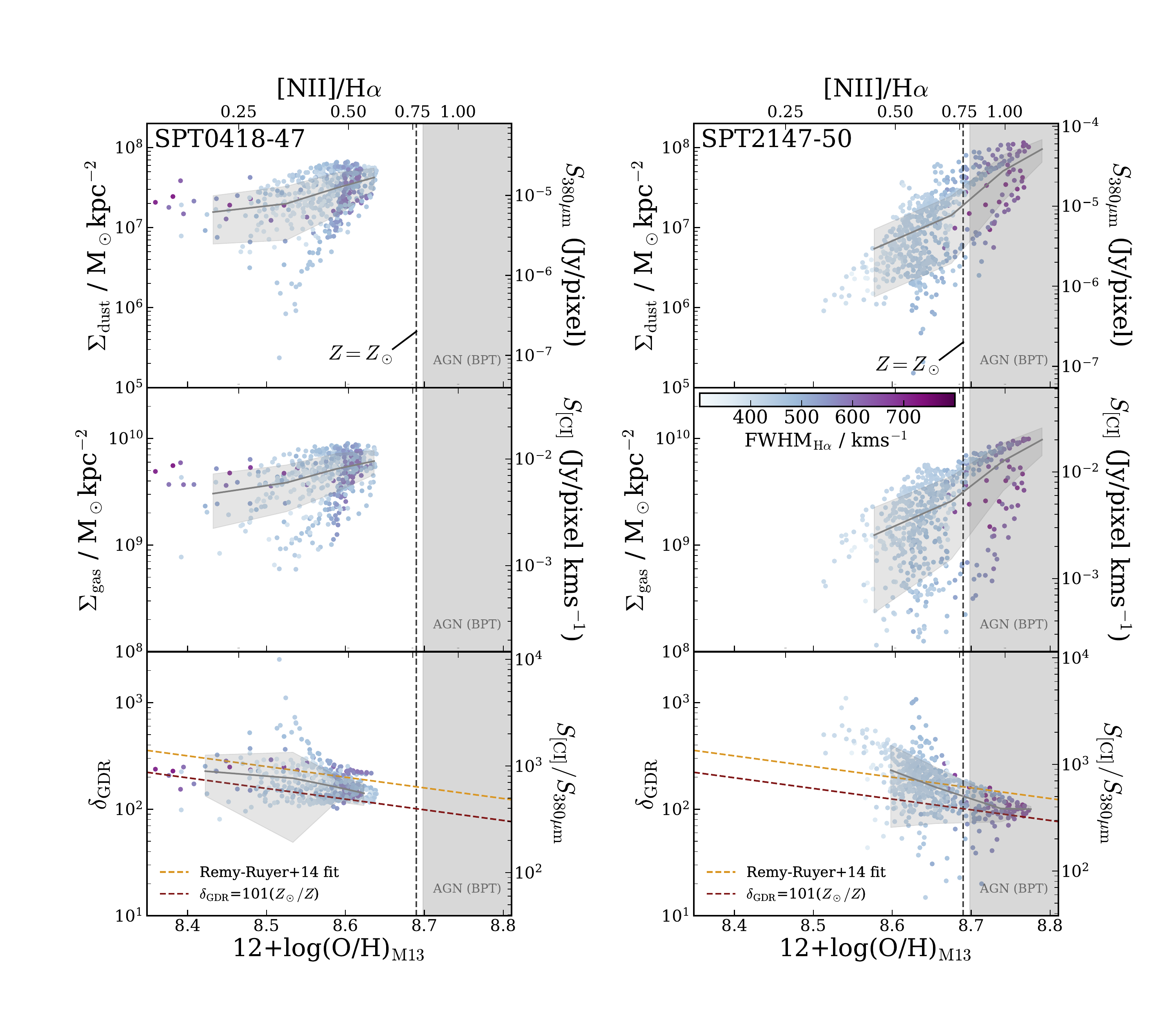}
\caption{Spaxel-by-spaxel (0.1$''$, $\sim$\,0.7\,kpc) variation of 380\,$\mu$m flux (top), [C{\sc i}] line flux (middle) and [C{\sc i}]/380\,$\mu$m flux ratio (bottom) as a function of [N{\sc ii}]/H$\alpha$ line ratio for the two SPT DSFGs. On the top axes we show the corresponding oxygen abundances according to the calibration of \protect\cite{marino13}. On the right axes we show the corresponding dust mass surface density $\Sigma_{\rm dust}$, gas mass surface density $\Sigma_{\rm gas}$, and gas-to-dust ratio $\delta_{\rm GDR}$, using the conversions described in \S\ref{sec:resolved_Z}. We plot a binned median (gray line) with 1-$\sigma$ scatter (gray region), and color-code points by H$\alpha$ FWHM. We also indicate the solar metallicity \citep{asplund21} and shade the region where [N{\sc ii}]/H$\alpha$\,$>$\,0.8, the latter of which is likely to be contaminated to some extent by AGN emission. Notably, in this region H$\alpha$ emission is broadest when compared to the rest of the source. In both sources the gas-to-dust ratio appears to decline with increasing metallicity, although this trend is more prominent in SPT2147-50. In the bottom panels we also show models from \cite{remy-ruyer14} and \cite{aniano20}, which are discussed further in \S\ref{sec:resolved_Z}.
}
\label{fig:gdr_Z}
\end{figure*}

%
%
\subsection{Resolved metallicities and the gas-to-dust ratio}
\label{sec:resolved_Z}

The impressive sensitivity and resolution of JWST has allowed us to spatially resolve the [N{\sc ii}]/H$\alpha$ flux ratio (see Figs.~\ref{fig:dust_gas_metal_maps_spt0418} and \ref{fig:dust_gas_metal_maps_spt2147}). We wish to compare these with the distribution of dust and gas in the two galaxies in order to determine how they are related. We therefore compare the [N{\sc ii}]/H$\alpha$ maps from NIRSpec with existing ALMA imaging of the rest-frame 380\,$\mu$m continuum and [C{\sc i}] line emission, which provide estimates of the dust and ISM masses.

Figs.~\ref{fig:dust_gas_metal_maps_spt0418} and \ref{fig:dust_gas_metal_maps_spt2147} show ALMA rest-frame 380\,$\mu$m continuum maps and [C{\sc i}] emission line maps of the two DSFGs, the latter of which are derived by summing the flux within twice the FWHM of the line. We also show the ratio of the [C{\sc i}] to 380\,$\mu$m maps, which we use as a proxy for the gas-to-dust ratio ($\delta_{\rm GDR}$). Visually, these three maps are broadly consistent in terms of regions of high [N{\sc ii}]/H$\alpha$ coinciding with regions of high flux in the ALMA maps. This is most clear in SPT2147-50, which displays two bright regions in the 380\,$\mu$m maps that are also present in the NIRSpec data.

To quantify the relation between the metallicity, gas, and dust, we convert the ALMA maps to dust and gas mass surface density $\Sigma_{\rm dust}$ and $\Sigma_{\rm gas}$ and plot both versus [N{\sc ii}]/H$\alpha$ on a pixel-by-pixel basis, as shown in Fig.~\ref{fig:gdr_Z}. To convert the ALMA rest-frame 380\,$\mu$m continuum maps to dust mass maps we apply Eq.~(1) of \cite{dunne03}, assuming a dust mass opacity coefficient at 125\,$\mu$m of 2.64\,m$^2$\,kg$^{-1}$, a dust temperature $T_{\rm dust}$ of 40\,K, and extrapolating to 380\,$\mu$m with dust emissivity index $\beta$\,$=$\,2.0. 

This approach ignores any potential variations in $T_{\rm dust}$ and $\beta$. However, \cite{spilker23} recently analysed rest-frame 120\,$\mu$m and 160\,$\mu$m data from SPT0418-47 and found no evidence of significant variations in dust temperature across the source. If indeed we are seeing AGN emission in SPT2147-50 then we may expect these regions to display higher dust temperatures, but we flag regions of the parameter space in Fig.\,\ref{fig:gdr_Z} where this is likely to be the case and avoid overinterpreting these regions. There has yet to be significant study of variations in the dust emissivity across individual galaxies, and indeed we cannot rule this out in our sources.

To convert the [C{\sc i}] emission line maps to molecular gas mass maps we use Eq.~(6) of \cite{bothwell17}, except in our case we use the Einstein $A_{21}$\,$=$\,2.68\,$\times$\,10$^{-7}$\,s$^{-1}$ coefficient \citep{papadopoulos04} and excitation factor $Q_{21}$\,$=$\,0.22 \citep{dunne22} for the [C{\sc i}](2--1) transition instead of the [C{\sc i}](1--0) values. Following \cite{papadopoulos04} and \cite{bothwell17} we adopt a [C{\sc i}]/H$_2$ abundance ratio of $X_{\rm [CI]}$\,$=$\,3\,$\times$\,10$^{-5}$. This is a commonly adopted value in the literature, but we note that this value cannot be measured for high-$z$ galaxies and is therefore based on measurements from local galaxies. It is therefore subject to considerable uncertainty. Indeed, through a comparison with dust-based gas masses, \cite{bothwell17} suggested that a higher value of $X_{\rm [CI]}$\,$=$\,7\,$\times$\,10$^{-5}$ could also be reasonable. Given that we are primarily concerned with the {\it variation} of $\delta_{\rm GDR}$ with metallicity, rather than its absolute value, this assumption should not significantly affect our conclusions.

In Fig.~\ref{fig:gdr_Z} we plot three quantities as a function of [N{\sc ii}]/H$\alpha$, a proxy for metallicity. These are $S_{380_{\mu{\rm m}}}$, $S_{\rm [CI]}$ and $S_{\rm [CI]}$/$S_{380_{\mu{\rm m}}}$, which we use as proxies for dust mass surface density, gas mass surface density, and gas-to-dust ratio respectively. The points are color-coded by H$\alpha$ FWHM. Given the various caveats and assumptions which come with these proxies, we choose to plot both the observed and converted quantities. Our interpretation will mostly focus on the variation of these properties rather than their absolute values. Additionally, all corresponding panels are shown on the same scale (of absolute values) to enable a direct comparison between the two galaxies. We also indicate the solar metallicity and shade the region where [N{\sc ii}]/H$\alpha$\,$>$\,0.8 which the BPT diagram suggests is likely to result from AGN emission.

Firstly, we note that in both galaxies, there appears to be a positive correlation between dust/gas mass surface density and metallicity. This indicates that regions containing more gas and dust are also more metal rich. This trend is stronger in SPT2147-50, in which we see that the most metal-rich regions also display the broadest H$\alpha$ emission. Given that these points fall within the shaded region, it is likely that the most ``metal-rich'' regions of SPT2147-50 are in fact regions of AGN emission. This is a striking conclusion given that no clear AGN have previously been identified in any SPT-selected DSFG, including some objects more extreme than those targeted here \citep[e.g.][]{ma16,apostolovski19}. It is theorized that DSFGs proceed to evolve through a quasi-stellar object (QSO) phase \citep{blain02,swinbank06, hopkins08}, and therefore we may be observing the beginnings of this evolution in SPT2147-50. NIRSpec follow up should be proposed for other sources in the SPT DSFG sample, in order to identify any further AGN candidates.

Turning to the bottom panels, we see that the gas-to-dust ratio appears to decline with increasing metallicity, a result that has been observed in local galaxies \citep[e.g.][]{leroy11,remy-ruyer14,aniano20}. In the bottom panels of Fig.~\ref{fig:gdr_Z} we plot two models to compare with our data. We show the model fit by \cite{remy-ruyer14} to 126 local galaxies spanning $\sim$\,2 dex in metallicity, derived from strong-line calibrations. Next we include a simple model from \cite{aniano20} in which the gas-to-dust ratio is estimated assuming that essentially all heavy elements are in dust grains, with solar relative abundances. This results in a linear relation between metallicity and $\delta_{\rm GDR}$. Both models appear to match the results from both galaxies within the uncertainties. Even if we ignore the absolute values, the trend of decreasing $\delta_{\rm GDR}$ with metallicity is generally well described by both models. We tentatively conclude that the gas-to-dust ratio and its variation with metallicity in these two DSFGs is consistent with findings from local galaxies, which is not necessarily unexpected but has not yet been shown for DSFGs as distant as the two we present in this paper.

%
%
\section{Conclusions}
\label{sec:conclusions}

We have presented two of the first spatially resolved metallicity maps of dust-obscured star-forming galaxies at $z$\,$\sim$\,4, utilizing IFU observations with JWST/NIRSpec. Both sources are detected in H$\alpha$ and [N{\sc ii}] at very high S/N, enabling us to perform resolved spectroscopy on these lensed systems. We find both SPT0418-47 and SPT2147-50 to be enriched to near-solar metallicity (conservatively $\sim$\,0.7\,$Z_\odot$), with evidence for AGN-like [N{\sc ii}]/H$\alpha$ ratios and broad H$\alpha$ emission. The derived oxygen abundances are consistent with the fundamental metallicity relation for both galaxies. Through a direct comparison with ALMA rest-frame 380\,$\mu$m continuum and [C{\sc i}] line maps at matched resolution, which we use as proxies for dust and gas mass, we find that the gas-to-dust ratio is negatively correlated with the metallicity, a result that is consistent with literature studies of local galaxies. At least qualitatively, it appears that these early-universe dusty galaxies bear some resemblance to expectations from $z$\,$\sim$\,0 galaxies, even on highly resolved spatial scales.

In the absence of [O{\sc iii}] and H$\beta$ coverage we interpret these results with caution, as the possibility remains that regions of high [N{\sc ii}]/H$\alpha$ are actually the result of AGN emission. Furthermore, detailed lens modelling analysis and source plane reconstruction of these data have not yet been carried out; this will be explored in future work. Regardless, this work presents the first resolved metallicity maps in DSFGs at $z$\,$\sim$\,4, and the data quality shows the potential of JWST to further our understanding of the processes that shape the ISM in high-redshift DSFGs.

\begin{acknowledgments}
TAH is supported by an appointment to the NASA Postdoctoral Program (NPP) at NASA Goddard Space Flight Center, administered by Oak Ridge Associated Universities under contract with NASA.
The SPT is supported by the NSF through grant OPP-1852617.
J.D.V. and K.P. acknowledge support from the US NSF under grants AST-1715213 and AST-1716127.
J.D.V. acknowledges support from an A. P. Sloan Foundation Fellowship.
JSS is supported by NASA Hubble Fellowship grant \#HF2-51446  awarded  by  the  Space  Telescope  Science  Institute,  which  is  operated  by  the  Association  of  Universities  for  Research  in  Astronomy,  Inc.,  for  NASA,  under  contract  NAS5-26555.
D.N. acknowledges support from the US NSF under grant 1715206 and Space Telescope Science Institute under grant AR-15043.0001 
M.A. acknowledges support from FONDECYT grant 1211951, ANID+PCI+INSTITUTO MAX PLANCK DE ASTRONOMIA MPG 190030, ANID+PCI+REDES 190194 and ANID BASAL project FB210003.
JWST is operated by the Space Telescope Science Institute under the management of the Association of Universities for Research in Astronomy, Inc., under NASA contract no. NAS 5-03127.
The National Radio Astronomy Observatory is a facility of the National Science Foundation operated under cooperative agreement by Associated Universities, Inc. 
This paper makes use of the following ALMA data: ADS/JAO.ALMA\#2018.1.01060.S and ADS/JAO.ALMA\#2021.1.00252.S.
ALMA is a partnership of ESO (representing its member states), NSF (USA) and NINS (Japan), together with NRC (Canada), MOST and ASIAA (Taiwan), and KASI (Republic of Korea), in cooperation with the Republic of Chile.
The Joint ALMA Observatory is operated by ESO, AUI/NRAO and NAOJ.

\end{acknowledgments}

\facilities{JWST, ALMA}
\software{{\tt astropy} \citep{astropy18}, CASA \citep{casa}}



\bibliography{bibliography}{}
\bibliographystyle{aasjournal}

\end{document}